\documentstyle[twocolumn,epsfig]{mn}
\newcommand{\mincir}{\raise
-2.truept\hbox{\rlap{\hbox{$\sim$}}\raise5.truept
\hbox{$<$}\ }}
\newcommand{\magcir}{\raise
-2.truept\hbox{\rlap{\hbox{$\sim$}}\raise5.truept
\hbox{$>$}\ }}
\newcommand{\minmag}{\raise-2.truept\hbox{\rlap{\hbox{$<$}}\raise
6.truept\hbox{$>$}\ }}
\newcommand{\be}{\begin{equation}}
\newcommand{\ee}{\end{equation}}
\newcommand{\ba}{\begin{eqnarray}}
\newcommand{\ea}{\end{eqnarray}}
\newcommand{\brr}{\begin{array}}

\newcommand{\err}{\end{array}}
\newcommand{\bc}{\begin{center}}
\newcommand{\ec}{\end{center}}

\title{Evolution of the Phase-Space Density of Dark Matter Halos
 and Mixing Effects in Merger Events}
\author[S. Peirani, F. Durier and J. A.\ de Freitas Pacheco]
{S\'ebastien Peirani, Fabrice Durier and Jos\'e A.\ de Freitas Pacheco
\\
Observatoire de la C\^ote d'Azur, B.P.4229, F-06304 Nice Cedex 4, France \\
emails: peirani@obs-nice.fr, durier@obs-nice.fr, pacheco@obs-nice.fr
}

\begin{document}

\maketitle

\begin{abstract}

Cosmological N-body simulations were performed to study the evolution of the phase-space density
$Q = \rho/\sigma^3$ of dark matter halos. No significant differences in the scale relations
$Q \propto \sigma^{-2.1}$ or $Q \propto M^{-0.82}$ are seen for ``cold" or ``warm" dark matter models.
The follow up of individual halos from $z = 10$ up to the present time indicate the existence
of two main evolutionary phases: an early and fast one ($10 > z > 6.5$), in which Q decreases
on the average by a factor of 40 as a consequence of the randomization of bulk motions and
a late and long one ($6.5 > z \geq 0$), in which Q decreases by a factor of 20 because of
mixing induced by merger events. The study of these halos has also evidenced that rapid and positive
variations of the velocity dispersion, induced by merger episods, are related to a fast decrease of
the phase density Q. 

\end{abstract}


\section{Introduction}

Presently, the large-scale structure in the galaxy distribution on scales 0.02 $< k <$ 0.15h Mpc$^{-1}$
is successfully explained in the cold dark matter (CDM) paradigm (Cole et al. 2005
and references therein). The CDM power spectrum on these scales derived from 
large redshift surveys as, for instance, the Anglo-Australian 2-degree Field 
Galaxy Redshift Survey (2dFGRS), is also consistent with the Lyman-$\alpha$ forest data in the 
redshift range 2$< z < $4 (Croft et al. 2002; Viel et al. 2003; Viel, Haehnelt \& Springel 
2004).

In spite of these impressive successes, there are still discrepancies between simulations and 
observations at scales $\leq$ 1 Mpc, which have been extensively discussed in 
the literature in the past years. In particular, the sharp central density cusp in dark matter halos, predicted
by simulations and not seen in the rotation curve of bright spiral galaxies
(Palunas \& Williams 2000; de Blok et al. 2001). 
We could also mention the large number of subhalos present in 
simulations but not observed (Kauffmann, White \& Guiderdoni 1993; Moore et al. 1999; Klypin 
et al. 1999), as in the case of our 
Galaxy or M31. Moreover, deep surveys (z$\geq$ 1-2) as the Las Campanas Infrared Survey, HST 
Deep Field North and Gemini Deep Deep Survey (GDDS) are revealing an excess of massive galaxies 
with respect to predictions of the hierarchical scenario (Glazebrook et al. 2004).

Massive halos are formed by continuous accretion and/or by merging. After a 
merger episode, the resulting halo is not in equilibrium, but after few dynamical time scales 
($t_{dyn} \sim (G\bar \rho)^{-1/2}$) the gravitational potential becomes almost steady. 
Violent relaxation produces a more mixed system (Tremaine, H\'enon \& Lynden-Bell 1986), reducing 
the value of the coarse-grained distribution function (DF), for which Q (defined as the ratio
between the density and the cube of the 1-D velocity dispersion in a given volume, e.g.,
$Q = \rho/\sigma^3$) is an estimator. Analyses of the core phase-space 
density of dwarf spheroidal galaxies, rotating dwarfs, Low Surface Brightness (LSB) galaxies and 
clusters of galaxies suggest that 
$Q\propto \sigma^{-n}$, with n $\sim$ 3-4 (Dalcanton \& Hogan 2001).
This behavior can be understood if the merging halos were initially almost in 
equilibrium and the fusion process preserves approximately the physical density as each layer is 
homologously added to form the new system (Dalcanton \& Hogan 2001). A scaling relation 
close to $Q\propto \sigma^{-3}$ was obtained from cosmological simulations by Dav\'e et al. (2001)
for collisionless dark matter as well as for self-interacting dark matter 
(SIDM). The result for SIDM is unexpected since in this case the material should be compressed to 
higher densities during the merger event, sinking to where it reaches the local pressure equilibrium and 
where the specific entropy matches. Therefore the merger should occur at nearly constant Q 
rather than constant density. Besides such a scaling, high resolution simulations of
galaxy-size CDM halos indicate an {\it increase} of the coarse-grained phase-
space density Q towards the center (Taylor \& Navarro 2001). Similar results were obtained 
by Rasia, Tormen \& Moscardini (2004), who have also obtained a 
power law variation, e.g., $Q \propto r^{-\beta}$ for cluster-size halos, with $\beta$ quite
close to the value found by Taylor \& Navarro (2001), namely, $\beta \approx$ 1.87. These
results suggest that the density profiles of dark matter halos are a consequence of a hierarchical
assembly process that preserves the phase-space stratification (Taylor \& Navarro 2001) or even
a generic feature of violent relaxation (Williams et al. 2004). 

In the present paper we report new cosmological simulations aiming to study the evolution
of the ``poor man's" phase-space density $Q$ in the core of dark matter halos. In particular, we
focus on the role of accretion/merger processes, which affect the dynamical relaxation of
halos, leading to structures with higher entropy or, equivalently, with lower phase-density Q.
We show that the Q-$\sigma$ scaling independs practically on the mechanism by which halos
acquire mass. Notwithstanding, a small differencial effect seems to exist, since halos which
have undergone major merger events have slightly lower phase-space densities than halos of
same mass, which have accreted matter quietly. The same behavior is noticed if warm dark
matter is considered instead of CDM. The second part of this investigation concerns the
follow up of some individual halos in order to obtain details of changes in the
phase-space density, occurring during merging events. We have found an early rapid decrease
of the phase-density Q, during and just after the first shell crossing, followed by a slow
and long decrease as halos accrete mass. Particles of captured subhalos diffuse in the
velocity space and after few dynamical timescales, share with particles of the main halo a
common (Gaussian) velocity distribution. This paper is organized as follows: in Section 2 the N-body 
simulations are brifly described, in Section 3 the evolution of the phase-space density Q as well as 
the effects of merger events are discussed and, finally, in Section 4 the main conclusions are given.

\section{Numerical simulations }

In this work, we use the same halo catalogs as in Peirani, Mohayaee \& de Freitas Pacheco (2004,
hereafter PMP04), but in the present study objects 
with masses higher than 10$^{13}$ M$_{\odot}$ were also included. For the sake of completeness, we
summarize here the main steps performed to prepare these catalogs.

The N-body simulation uses the adaptive particle-particle/particle-mesh (AP$^3$M) code HYDRA
(Couchman, Thomas \& Pearce 1995). The adopted cosmological parameters were h = 0.65, $\Omega_m$ = 0.3
and $\Omega_{\Lambda}$ = 0.7, with the power spectrum normalization $\sigma_8$ = 0.9. The simulation
was performed in a box of side 30h$^{-1}$ Mpc including 256$^3$ particles, corresponding to
a mass resolution of $2.05 \times 10^8 \, M_{\odot}$. The simulation started at z = 50 and ended at
the present time (z = 0). A similar simulation (here referred as S2) was also performed for 
a ``warm" dark matter (WDM) model. In this case, the 
fluctuation power spectrum becomes $P(k) = AkT_{cdm}^2(k)T_{wdm}^2(k)$, where the adopted transfer 
function for WDM is given by (Bode, Ostriker \& Turok 2001)
\begin{equation}
T_{wdm}(k) = \lbrack 1 + (\alpha\,k)^{2.4}\rbrack^{-25/6}
\end{equation}
where $\alpha$ = 0.102 for WDM particles having a mass of 0.5 keV and k is in hMpc$^{-1}$. This
particle mass corresponds to a free-streaming mass of about $1.7\times 10^{11}\, M_{\odot}$, scale
below which structures become to be suppressed. A third simulation (here referred as S3) with the 
same cosmological parameters but including 128$^3$ CDM particles was also run. However, particles were
separated into two 
categories, the first having masses ten times higher than the second. This simulation was carried out 
using the code GADGET (Springel, Yoshida \& White 2001) and was used to study the energy transfer
between particles during merger events. Here, the ``softening" length was set equal to 12 h$^{-1}$kpc.

Halos were initially detected by using a friends-of-friends (FOF) algorithm
and, in a second step, unbound particles were removed by an iterative procedure. Thus, 
halos of both catalogs are all gravitationally bound objects. For further details,
the reader is referred to PMP04.

The {\it accretion} catalog, comprising 781 objects, includes halos which have never undergone a 
major merger event and whose masses varied continuously and smoothly. 
The {\it merger} catalog contains 567 halos which had at least one major merger
episode in their history, corresponding to an increase of their masses at least by a factor of
1/3 in the event. In latter class, violent variations of the gravitational potential occur while
in the former category the potential varies smmothly all the time.
Halos were followed from z = 3.5 until z = 0, but about 40 objects, with
enough particles (N $\geq$ 50), were followed during a longer time interval (from z = 10 until z = 0).
From the simulation S2, 216 halos were included in the {\it merger} catalog and 334 halos in the
{\it accretion} catalog (3.5$\geq z \geq$ 0) but only 5 halos have been followed in the interval
$10 \geq z \geq 0$.

\begin{figure}

  \begin{center}
    \rotatebox{0}{ \includegraphics[height=6cm,width=8cm]{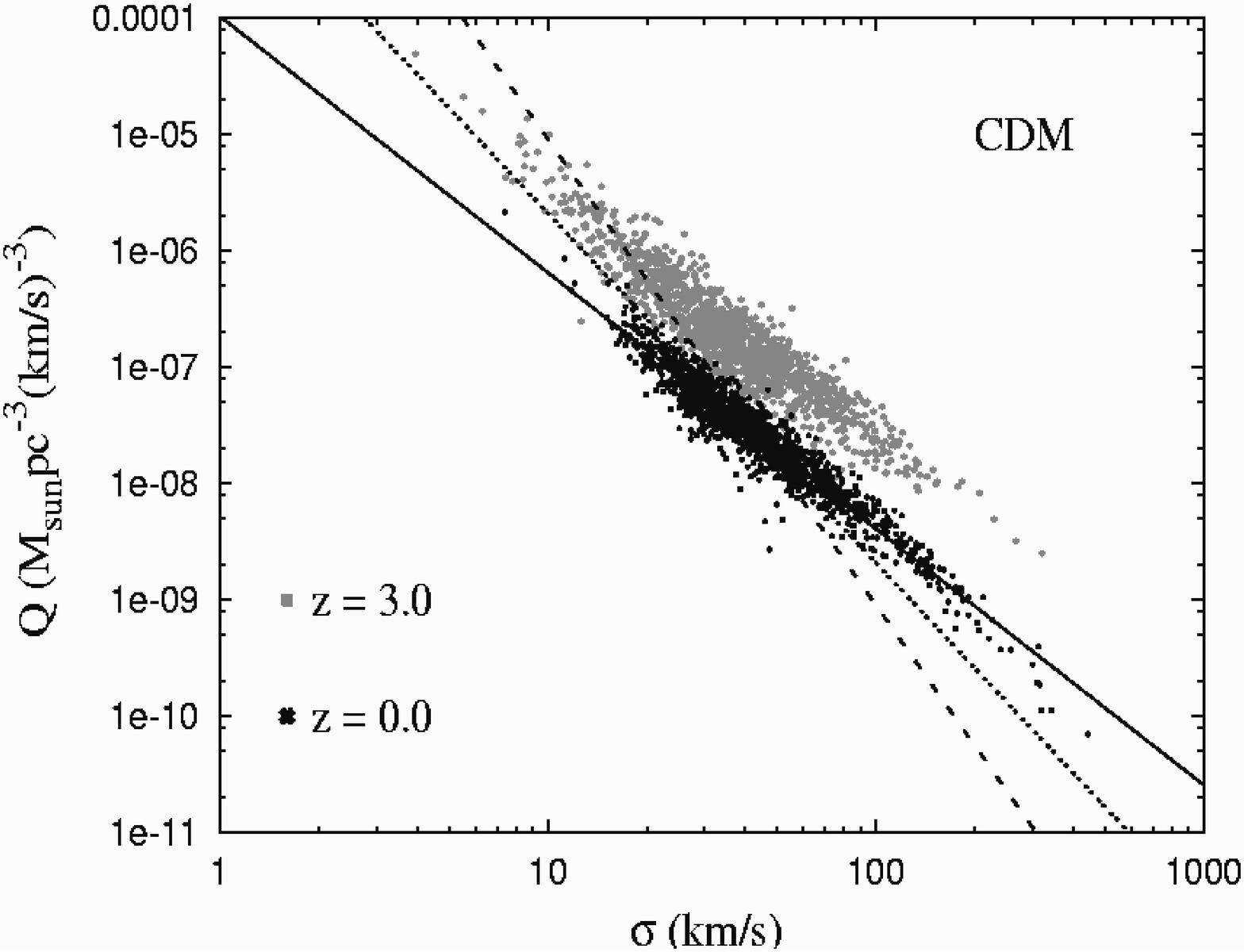}}
  \end{center}
  \vfill
  \vspace{-8mm}
  \begin{center}
    \rotatebox{0}{ \includegraphics[height=6cm,width=8cm]{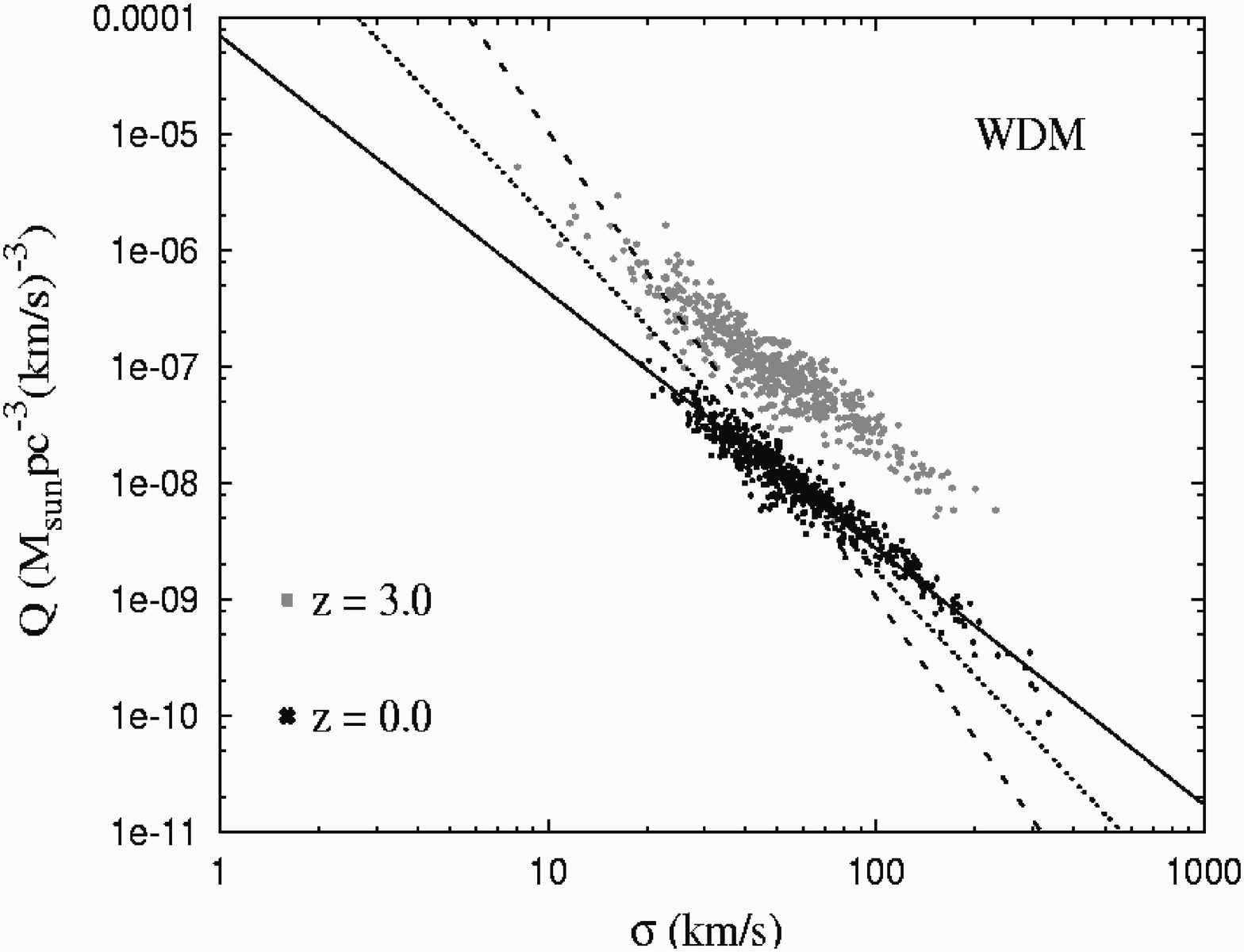}}
  \end{center}
\caption
{The phase-space density Q for halos of both catalogs as a function
of the 1D velocity dispersion $\sigma$ is shown for the CDM model (upper panel) and for
WDM model (lower panel). In each panel, the solid line indicates the best
fit solution ($Q\propto \sigma^{2.2}$). For comparison, lines representing $Q\propto \sigma ^{-3}$  (dotted)
and $Q\propto \sigma ^{-4}$ (dashed), normalized at the same point, were also included in the plot.}
\label{fig1}
\end{figure}

\begin{figure}
\centerline{
        \vspace{-0.8cm}
        \epsfxsize=0.5\textwidth\rotatebox{0}
        {\epsfbox{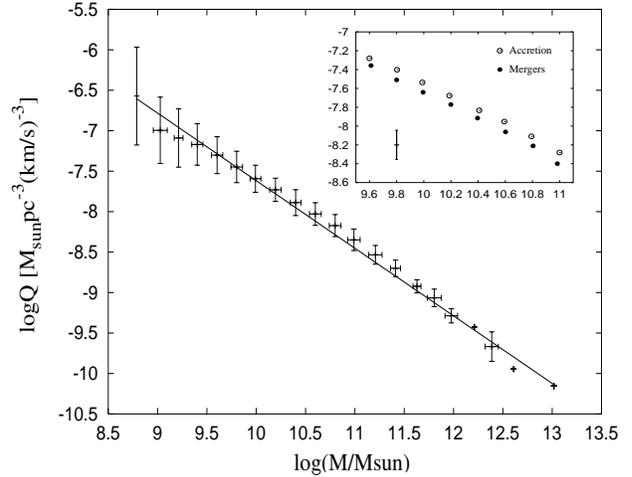}}
           }
\vspace{0.8cm}
\caption
{ The phase-density Q as a function of the core mass for halos (CDM model) of both catalogs
 at $z=0$. The solid line shows the best fit ($Q\propto m^{-0.83}$) and the small 
panel shows the same plot, but discriminates objects of different catalogs. The vertical bar indicates a
typical uncertainty in Q.}
\label{disc1}
\end{figure}

\section{The phase-space density evolution}

\subsection{Scaling relations}

One of the goals of this work is to study the evolution of the phase-space density $Q = \rho/\sigma^3$,
in order to verify how the ``violence" or the ``quietness" of the accretion process may affect the
relaxation road to equilibrium and the resulting scaling relations $Q \propto \sigma^{-n}$ or
$Q \propto M^{-k}$ .

For each halo at a given redshift, the value of Q was computed by adopting the following procedure.
Firstly, the halo center is defined as being coincident with the position of the most bound particle, since
during a merger event the center of mass is ``ill-defined".
In a second step, all particles inside a radius r$_*$ around the center are selected and r$_*$ is taken to be
equal to one tenth of the gravitational radius, e.g., r$_*$ = 0.1GM$^2$/$\mid W\mid$, where W is
the total gravitational energy of the system. This radius was chosen in order to include enough
particles necessary to compute adequately the density as well as the velocity dispersion and 
still be representative of the core region.
Instead of evaluating the core density by a simple 
mean ($\bar{\rho_c} = 3mN_c/4\pi r_*^3$), we have estimated the density $\rho_i$ at the position of each 
selected particle {\bf i} by using the Voronoi tessellation procedure in order to increase the accuracy of
the density estimation when the number of particles is small. In this technique, the
3-D space is divided into polyhedral cells centered on each particle, with boundaries defined 
in order to include all points closer to the center (see, for instance, van de Weygaert
1994, for details). The
cell density is simply the mass of the particle divided by the volume of its cell. Calculations
were made with the help of the public package QHULL\footnote{http://www.geom.umn.edu/software/qhull/}.
Finally, the core density of the halo is estimated from 
\begin{equation}
\rho_c = (1/N_c)\sum^{N_c}_i \rho_i
\end{equation}
and the 1-D velocity dispersion from $\sigma = \sqrt{(\sigma_x^2 + \sigma^2_y + \sigma^2_z)/3}$.

In Fig. 1, the phase-space density Q for all halos and for both CDM and WDM models is shown as a function
of the 1-D velocity dispersion $\sigma$ at $z=3.0$ and $z=0$. 
For the CDM model, a best fit to the data
indicates that $Q\propto \sigma^{-2.08}$ for $z=3.0$ and $Q\propto \sigma^{-2.16}$ for $z=0$.
No significant modifications in the exponent n ($Q \propto \sigma^{-n}$) are noticed if the definition of
$r_*$ is changed. For instance, at $z = 0$, n = 2.23 and n = 2.18 if $r_*$ is decreased by a factor
of two or increased by 50\% respectively. 
We have also investigated how our results could be affected by using the mean density $\bar{\rho_c}$
instead of the tessellation method. In fact, a slight increase in the exponent n is observed (n = 2.64
at $z = 3$ and n = 2.53 at $z = 0$). The reason for such differences is the following: the density
evaluated from the Voronoi procedure in a given volume is weighted by the inverse cell volume associated
to each considered particle. Thus, if important density gradients are present, as those observed
in DM halos, the two density estimations differ. When using simulated data, this effect in more
pronounced when massive halos are considered, since the number of particles is large, strengthening
the gradient effect. As a consequence, the use of a simple mean underestimates the density as well as
the phase-density $Q$ with respect to the Voronoi method. This effect is minimized for low mass halos,
since the number of particles inside the considered volume is small, diluting the gradient effect.
In order to give some numbers, the core volume of a halo of $7\times 10^{12}\, M_{\odot}$ including
$\sim$ 3200 particles is comparable
to the total cell volume, but the density derived by the tessellation procedure is about a factor of
two higher than the simple mean density. However, both density estimations give similar results if
a typical halo of $1.2\times 10^{11}\, M_{\odot}$, including $\sim$ 55 particles, is considered.
Some differences can also be seen when our simulated data is compared with observations
(see, for instance, Fig. 1 in Dalcanton \& Hogan 2001), since $Q$ values estimated for
low brightness galaxies are higher than our expectations. This is probably a consequence that density
estimations from observational data include the contribution of baryonic matter, dominant in the central 
regions of these objects (Ortega \& de Freitas Pacheco 1993), which is not the case for clusters. 

It is worth mentioning that the exponent {\it n} found from our simulations can be understood on the basis 
of the following considerations: from data of our catalogs, halos acquire mass continuously and evolve according 
to the scaling relation M $\propto \sigma^{8/3}$. If they evolve {\it almost} in virial equilibrium in the
considered redshift interval, then Q $\propto \sigma^{-7/3}$, in good agreement with our results.
It should also be emphasized that by the time of halo formation, the particle thermal velocities
are quite small and the early random motions, present when simulations begin, are a consequence
of gravitational ``heating" by primordial density fluctuations. This is true whether cold or warm
dark matter is considered. Thus, if scaling relations can be checked by simulations, this is not the case 
for the normalization of the initial phase-space density. Moreover, as halos evolve from z = 3 to z = 0,
the Q-$\sigma$ relation becomes slightly but significantly steeper and,
at a given velocity dispersion, the value of Q decreases by a factor of 4 in the same redshift
interval. A recent perturbative analysis of the spherical gravitational
collapse (Amarzguioui \& Gro$\phi$n 2005) indicates that the entropy increases (e.g., the phase-space $Q$ decreases)
in the process, as a consequence of potential energy conversion into random kinetic motions. This is in agreement 
with our findings, which covers the non-linear regime.

Halos in a WDM model have, on the average, systematically
higher virial ratios ($2T/\mid W\mid$) than CDM halos, i.e., they are less relaxed. In spite of this, no 
significant differences are seen in the $Q-\sigma$ diagram between both dark matter models. The exponent 
is equal to n = 1.96 ($z = 3.0$) and n = 2.13 ($z = 0$), comparable to the values found for the CDM model.

The evolution of Q as a function of the core mass of the halo for objects included in the {\it accretion} and
in the {\it merger} catalogs is shown in Fig. 2, for $z=0$ in the case of CDM (similar results are
obtained for WDM). Data were bined in logarithm intervals
of $\Delta log(M/M_{\odot}) = 0.20$ for a higher clarity of the plot.  Best fit solutions give
$Q\propto M^{-0.80}$ for the accretion and $Q\propto M^{-0.82}$ for the merger samples respectively.
Again, using the empirical scaling relation derived from our simulated data, e.g., $M \propto \sigma^{8/3}$
and the assumption that halos evolve in quasi-equilibrium from $z = 3.0$ up today, one obtains
$M \propto \sigma^{-7/8}$, quite comparable to the values derived from simulated data.

\subsection{Study of individual halos}

Halos are continuously accreting mass (violently or not). This process affects their dynamical state and
contributes to decrease their phase-space density $Q$.
Thus, scaling relations as  $Q \propto \sigma^{-n}$ or $Q \propto M^{-k}$  should be considered 
in a statistical sense and they are not expected to describe the evolution of any particular 
halo, whose evolution depends on his accretion history.

In order to understand the early behavior of the coarse-grained DF (measured by Q),
43 CDM and 5 WDM halos in the mass range M $\sim$ 10$^{11-12}$ M$_{\odot}$ were followed since z = 10 up today.
Criteria to identify a given halo at different redshifts were discussed in our previous work (PMP04).
For each halo, we have evaluated the phase-space density Q, the core density $\rho_c$ and the velocity
dispersion $\sigma$ as a function of time. Some typical examples are exhibited in Fig. 3, where the 
evolution of these parameters for 3 individual halos (2 CDM and one WDM) is shown.

\begin{figure*}
\includegraphics[width=16.0cm]{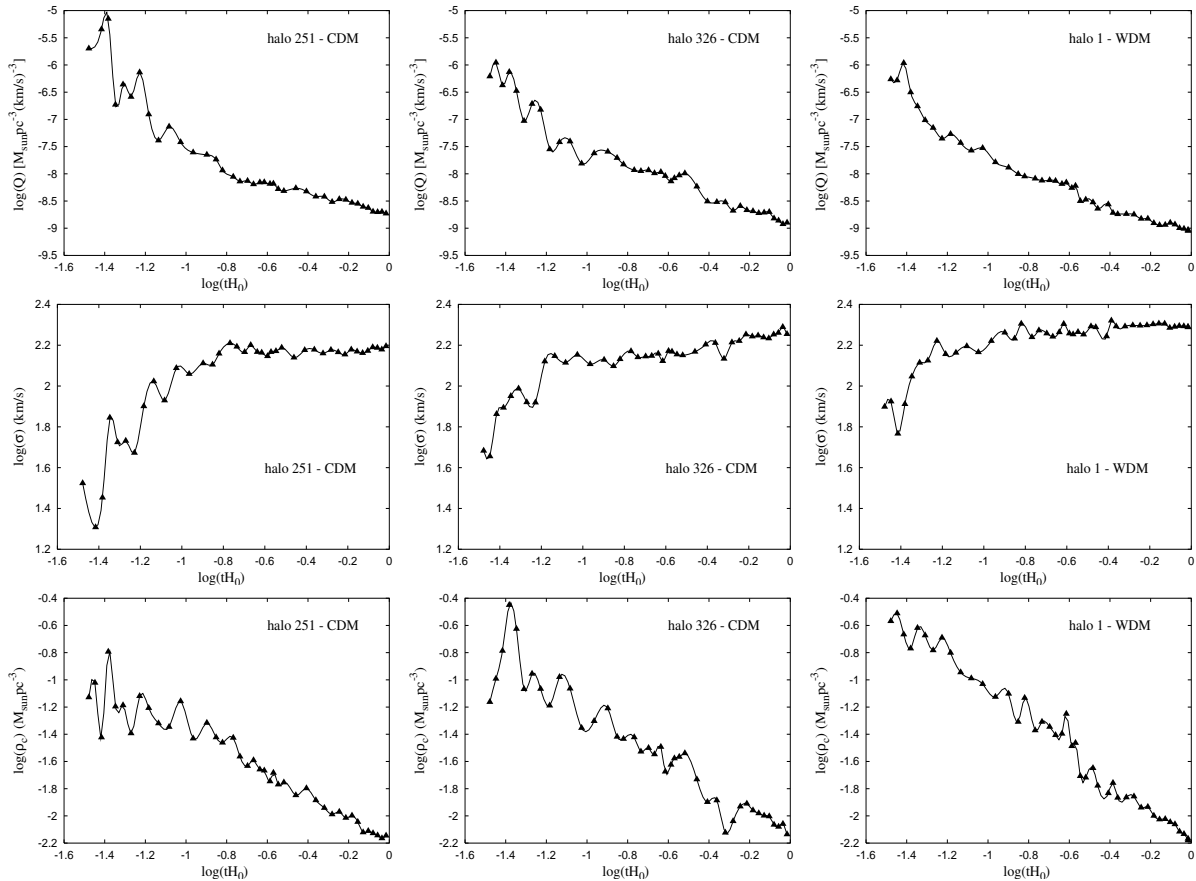}
\caption{The evolutions of Q, $\sigma$ and $\rho_c$ for 3 halos from $z=10$ to $z=0$. Panels in
the first two columns concern halos evolving in a CDM model while panels in the last column
concerns an example of a halo evolving in a WDM model.}
 \label{figure3}
 \end{figure*}


\begin{figure*}
\includegraphics[width=14.0cm]{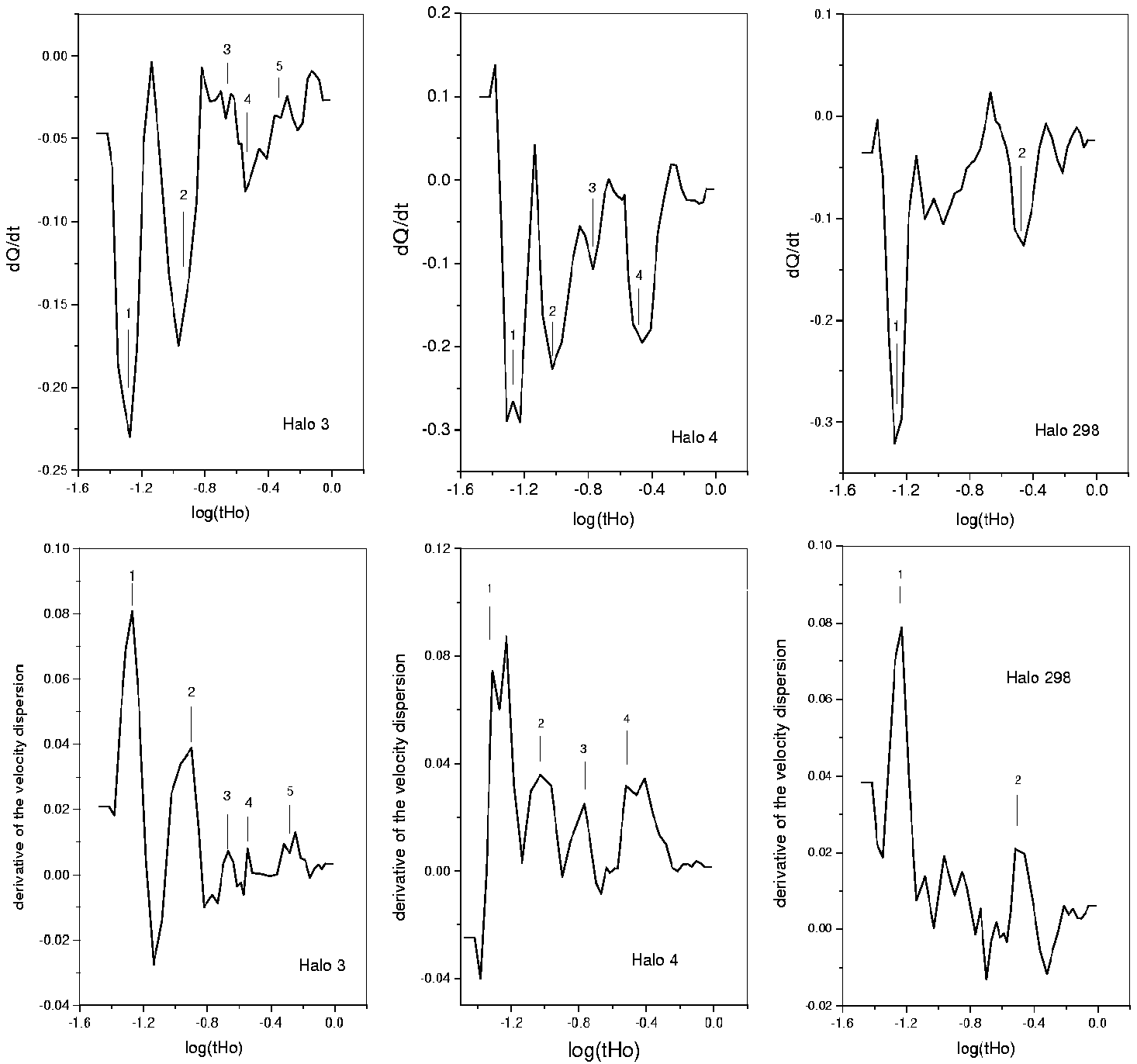}
\caption{ The evolution of the time derivative of Q and those of the velocity dispersion for 3 examples of 
CDM halos. The first event labelled 1 corresponds in general to the randomization of bulk motions, while
the others correspond to merger events.}
 \label{figure4}
\end{figure*}


As halos accrete mass, they approach a state of dynamical equilibrium, measured by their
virial ratio 2T/$\mid W\mid$, which approaches asymptotically
the unity. On the one hand, their gravitational radius increases continuously and, as 
a consequence, the core density $\rho_c$ decreases on the average by one order of magnitude 
in the interval 10$\leq z \leq$ 0, for halos in the considered mass range, indicating that
the evolution of Q can not be derived by a simple homologous model. On the other hand,
from z = 10 up to about z $\sim$ 6.5, there is a rapid increase of the velocity dispersion, associated
to the process of relaxation, followed by a phase of slow ``heating". In the early evolutionary
phases, an important energy transfer from bulk to random motions occurs due to collective 
effects, heating all the particles, regardless their initial energies. 
Such a transfer by collective effects is clearly seen in merger episodes issued from simulation
S3, indicating that particles with different masses have the same final dispersion velocity.
This relaxation process reduces considerably the phase-space density in a short time
interval. Q decreases, on the average, by
a factor of 40 in a time interval of about 0.5 Gyr, whereas in the next phase, which lasts about
13 Gyr, the phase-space density decreases by factor of 20 due to  phase-mixing mechanisms
induced by mass accretion processes.

The effect of the ``heating" in variations of the phase-space density can be clearly seen if
the time derivatives of Q and $\sigma$ are compared. The time derivative of the velocity dispersion 
has, in general, clear peaks indicating phases of
fast ``heating". The first and principal peak is associated to the ``randomization" of initial
bulk motions and the others are associated to merger events. The amplitude of these secondary peaks
depends on the mass of the captured subhalo. These events produce a sudden decrease of
the coarse-grained DF and a more relaxed system, with a higher entropy. In Fig. 4 we show three examples
of  CDM halos where effects of merger events seen in the evolution of the velocity dispersion are
clearly anti-correlated with phase-space density variations.

\section{Conclusions}

The dynamical evolution of dark matter halos depends on their accretion history. During merger
events strong mixing effects occur away from an equilibrium state, driven by large-scale fluctuations
of the gravitational potential. In this work we have investigated different aspects of mixing
through analyses of the phase-space density Q evolution after
accretion/merger events.

Our simulations indicate that for $z \leq 3.0$ the phase-space density scales as $Q \propto \sigma^{-2.1}$
and that such a scaling does not depend on the adopted dark matter model. The same behavior is
obtained either for cold or warm dark matter, in spite of CDM halos be always more relaxed, at
a given redshift, than WDM halos of comparable mass. From the scaling relation derived from our simulations
between the halo mass and the velocity dispersion, considering also that at those redshifts
halos have virial ratios near unity, we obtain 
$Q \propto \sigma^{7/3}$, consistent with the results derived directly from our numerical experiments.

Q values derived for halos which have underwent important merger events are, on the 
average, slightly lower than those derived for halos which have grown by accretion only. This is 
consistent with the
expectation that violent variations of the gravitational potential, which occur in a
merger event, lead to a more mixed system. Additional information about the evolution of 
the phase-space density $Q$ was obtained from the follow up of individual halos 
from $z \approx 10$ up today. In the early evolutionary phases
($10 > z > 6.5$), $Q$ decreases on the average by a factor of 40, as a consequence mainly of the
randomization process of the initial bulk motions. In a late (and long) phase ($6.5 > z \geq 0$),
$Q$ decreases by a factor of 20 on the average, consequence of continuous mass accretion. These
processes are well evidenced when the time derivative of the phase density $Q$ and that
of the velocity dispersion for individual halos are compared. There is a clear anticorrelation
between these two quantities: rapid {\it positive} variations of the velocity dispersion, induced 
by merger events, are connected with a fast decrease of the phase density $Q$.  

\vspace{1.0cm}

\noindent
{\bf Acknowledgment}

\noindent
S.\ P.\ acknowledges a PhD fellowship from Universit\'e de Nice Sophia-Antipolis (UNSA) and
F.\ D.\ acknowledges the Observatoire de la C\^ote d'Azur and the Conseil R\'egional PACA
for the support of his PhD program. We thank the referee for the useful comments which have
contributed to improve the text of this paper.
S.P. thanks  H. Mathis for his kindness and
instructive conversations. S.P. also thanks E. Lagadec for ``chouchenn".


\end{document}